\documentclass[11pt]{article}

\usepackage[preprint]{acl}

\usepackage{times}
\usepackage{latexsym}

\usepackage[T1]{fontenc}

\usepackage[utf8]{inputenc}

\usepackage{microtype}

\usepackage{inconsolata}

\usepackage{graphicx}

\usepackage{minitoc}
\usepackage{amsfonts}
\usepackage{amssymb}
\usepackage{amsmath}
\usepackage{algorithm}
\usepackage{algpseudocode}
\usepackage{booktabs}
\usepackage{listings}
\usepackage{multirow}
\usepackage{pifont} 
\usepackage{siunitx}
\newcommand{\xmark}{\ding{55}}
\lstdefinestyle{mystyle}{
    language=Python,
    basicstyle=\ttfamily\footnotesize,
    breaklines=true,
    showstringspaces=false,
    commentstyle=\color{green},
    keywordstyle=\color{blue},
    stringstyle=\color{red},
    backgroundcolor=\color{yellow!10}
}
\title{UrgentMOS: Unified Multi-Metric and Preference Learning for Robust Speech Quality Assessment}

\author{
Wei Wang$^1$,
Wangyou Zhang$^1$,
Chenda Li$^{1}$,
Jiahe Wang$^{1}$,
Samuele Cornell$^2$,
Marvin Sach$^3$, \\
\textbf{
Kohei Saijo$^5$,  
Yihui Fu$^3$,
Zhaoheng Ni$^4$, 
Bing Han$^1$,
Xun Gong$^1$,
Mengxiao Bi$^6$,
}  \\
\textbf{
Tim Fingscheidt$^3$,
Shinji Watanabe$^2$,
Yanmin Qian$^1$} \\
$^1$Shanghai Jiao Tong University, China 
$^2$Carnegie Mellon University, USA \\
$^3$Technische Universität Braunschweig, Germany
$^4$Meta, USA \\
$^5$Waseda University, Japan 
$^6$VUI Labs, Hangzhou, China \\
}

\begin{document}
\maketitle
\begin{abstract}
Automatic speech quality assessment has become increasingly important as modern speech generation systems continue to advance, while human listening tests remain costly, time-consuming, and difficult to scale. Most existing learning-based assessment models rely primarily on scarce human-annotated mean opinion score (MOS) data, which limits robustness and generalization, especially when training across heterogeneous datasets. In this work, we propose UrgentMOS, a unified speech quality assessment framework that jointly learns from diverse objective and perceptual quality metrics, while explicitly tolerating the absence of arbitrary subsets of metrics during training. By leveraging complementary quality facets under heterogeneous supervision, UrgentMOS enables effective utilization of partially annotated data and improves robustness when trained on large-scale, multi-source datasets. Beyond absolute score prediction, UrgentMOS explicitly models pairwise quality preferences by directly predicting comparative MOS (CMOS), making it well suited for preference-based evaluation scenarios commonly adopted in system benchmarking. Extensive experiments across a wide range of speech quality datasets, including simulated distortions, speech enhancement, and speech synthesis, demonstrate that UrgentMOS consistently achieves state-of-the-art performance in both absolute and comparative evaluation settings.
\end{abstract}

\section{Introduction}

Automatic speech quality assessment has become an essential component for evaluating generative speech AI systems such as speech synthesis, enhancement and coding~\cite{huang22f_interspeech,yi22b_interspeech,shi2024espnet}. This task has received increasing attention in recent years due to the rapid advancement of speech generation models and the growing demand for reliable and scalable evaluation methods. However, the gold standard for assessing speech quality remains human listening tests, which are costly, time-consuming, and difficult to scale~\cite{anastassiou2024seed,guo2025splitmeanflow,ditar-jia2025,F5TTS-chen2024}.
Evaluating modern speech generation systems with human judgments remains challenging beyond issues of cost and scalability. As synthesized speech approaches natural quality, perceptual differences between systems become increasingly subtle, making judgments more sensitive to listener bias and preference. Even under controlled protocols, listeners may prioritize different quality aspects—such as naturalness, intelligibility, or residual artifacts—leading to higher annotation noise and reduced consistency~\cite{zielinski2008some,loizou2011speech,jimenez2021removing,naderi2020towards,sach2025p}. These limitations are especially evident in Absolute Category Rating (ACR) settings, where score saturation and inter-listener variability weaken discriminative power. Consequently, there is growing interest in comparison-based evaluation paradigms that directly assess relative quality differences~\cite{SpeechEval,SpeechJudge}, which better support system-level comparisons and provide more stable judgments when absolute ratings become unreliable. This shift motivates speech quality assessment models that can natively learn from comparative supervision.

Objective speech quality assessment has been widely studied as a scalable alternative to human listening tests~\cite{MCD-Kubichek1993,PESQ-Rix2001,SDR-Vincent2006,STOI-Taal2011}. Early methods relied on signal processing and psychoacoustic principles, often requiring clean reference signals and capturing only limited aspects of speech quality, which restricts their applicability in real-world settings. More recently, neural Mean Opinion Score (MOS) predictors~\cite{NISQA-Mittag2021,DNSMOS-Reddy2022,SCOREQ-Ragano2024,URGENT-SQA,Distill_MOS-Stahl2025} have emerged to estimate perceptual quality without references. However, large-scale and consistent MOS annotations remain scarce, and MOS scores are often poorly aligned across datasets due to differences in annotation protocols and listener populations. As a result, models trained solely on MOS supervision can exhibit limited robustness, especially when learning from heterogeneous data.

Motivated by these challenges, we propose UrgentMOS, a unified speech quality assessment framework inspired by the URGENT Challenges~\cite{URGENT-Zhang2024, URGENT-Kohei2025, li2026icassp}, in which speech enhancement systems are evaluated using a diverse set of objective and perceptual metrics under both ACR and Comparison Category Rating (CCR) protocols. To address the instability and limited comparability of human MOS annotations across datasets, UrgentMOS adopts multi-metric supervision, grounding MOS prediction in multiple complementary quality facets. Robust generalization is achieved by training on a large and diverse collection of datasets while allowing arbitrary subsets of metrics to be absent, such as intrusive metrics that require reference signals or large-scale corpora without human MOS annotations. In addition, to meet the growing demand for comparative evaluation, UrgentMOS directly incorporates preference prediction into the training objective by explicitly modeling paired speech samples from preference-annotated datasets as well as carefully constructed pairs derived from ACR datasets, enabling unified support for both absolute and comparative speech quality evaluation.
Our contributions can be summarized as follows:
\begin{itemize}
\item We propose UrgentMOS, a unified speech quality assessment framework that jointly learns from objective and perceptual metrics, enabling robust training across heterogeneous datasets with incomplete annotations.
\item UrgentMOS aligns model training with practical evaluation protocols by directly supporting both absolute and comparative supervision.
\item We construct and release a preference annotated speech quality dataset derived from existing ACR datasets, providing a benchmark for comparative speech quality assessment.
\item Extensive experiments on diverse speech quality datasets show that UrgentMOS achieves state-of-the-art performance in both absolute and comparative evaluation settings.
\end{itemize}

\begin{figure*}[t] 
    \centering
    \includegraphics[width=\textwidth]{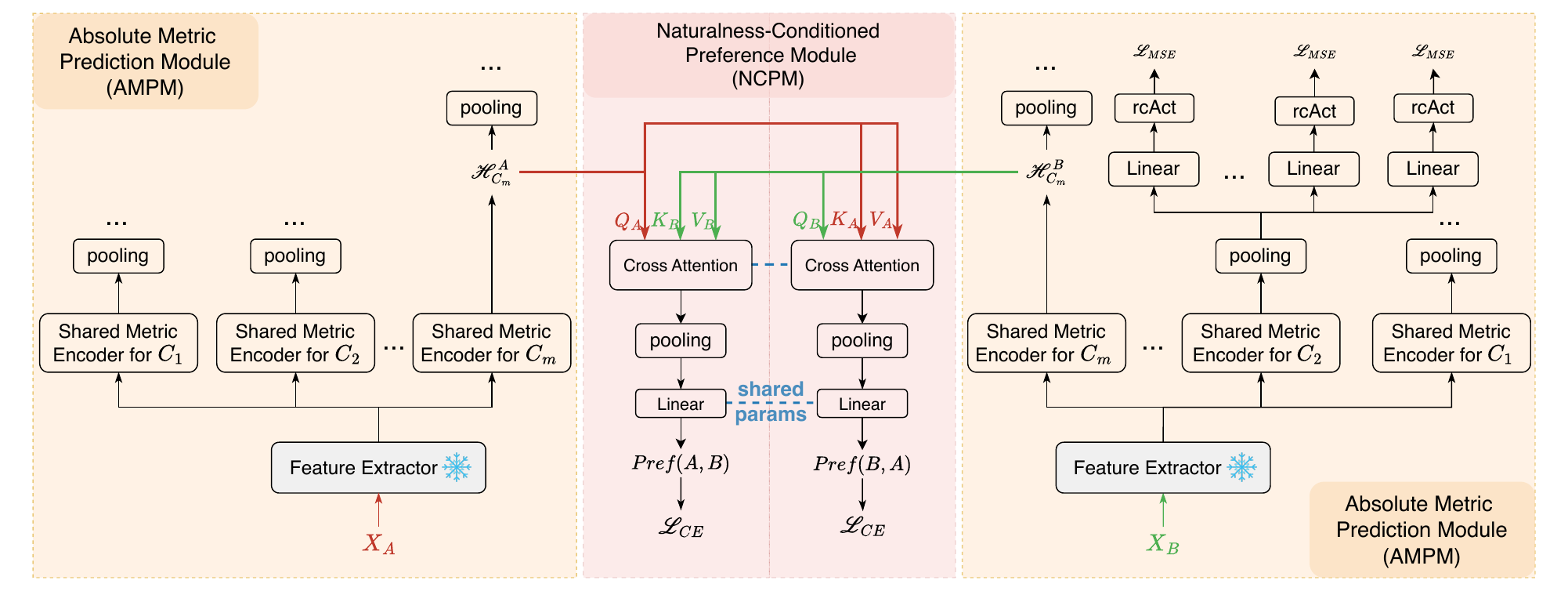} 
\caption{Overview of the UrgentMOS architecture and training paradigm.
UrgentMOS consists of an \textit{Absolute Metric Prediction Module (AMPM)} and a \textit{Naturalness-Conditioned Preference Module (NCPM)}.
Each speech sample is processed by a shared feature extractor; the two AMPM blocks shown for Audio A and Audio B correspond to the \emph{same AMPM with shared parameters}. AMPM predicts multiple objective and perceptual quality metrics via shared metric encoders and metric-specific heads, while NCPM operates on representations from the naturalness-related metric group to model pairwise quality preferences using cross-attention. $\mathcal{L}_{CE}$ denotes cross entropy loss. $\mathcal{L}_{MSE}$ is detailed in (\ref{eq:loss_mse}).}
\vspace{-1em}
    \label{im:overall}
\end{figure*}

\section{Related Works}

\subsection{Absolute Score based Speech Quality Assessment}

Learning-based speech quality assessment has primarily focused on predicting absolute Mean Opinion Scores (MOS) from speech signals. MOSNet~\cite{MOSNet-Lo2019} formulates MOS prediction as supervised regression using neural networks trained on human-annotated data. DNSMOS~\cite{DNSMOS-Reddy2022} introduces a multi-stage self-teaching framework for non-intrusive evaluation of noise suppression systems with strong correlation to human ratings. DNSMOS Pro~\cite{dnsmospro_cumlin24} extends DNSMOS with a lightweight architecture and probabilistic MOS prediction, enabling efficient training using only MOS labels across datasets. NISQA-MOS~\cite{NISQA-Mittag2021} employs self-attention--based temporal modeling to predict overall MOS and multiple perceptual dimensions. UTMOS~\cite{UTMOS-Saeki2022} and UTMOSv2~\cite{UTMOSv2-Baba2024} improve robustness to out-of-domain conditions through ensemble learning over self-supervised speech representations. SCOREQ~\cite{SCOREQ-Ragano2024} addresses domain generalization by introducing a contrastive regression objective with triplet loss for MOS prediction. Distill-MOS~\cite{Distill_MOS-Stahl2025} improves data efficiency and model compactness by distilling large self-supervised MOS predictors into smaller models using pseudo-labels on unlabeled data.  
However, in practical evaluation scenarios where model improvements across iterations are often marginal, absolute MOS prediction can be insensitive to small quality differences, making pairwise comparison a more suitable formulation for detecting subtle improvements by directly contrasting two speech samples.

\subsection{Preference based Speech Quality Assessment}

Preference-based speech quality assessment models relative quality differences between speech samples. Early work reformulates MOS prediction using pairwise comparisons, demonstrating improved ranking performance when absolute MOS values are close~\cite{wang2023mospc}. Subsequent studies encode preference relationships in embedding spaces to improve robustness under domain shifts~\cite{hu2024embedding_learning}, or propose end-to-end frameworks that jointly model pair construction and preference prediction for system-level ranking~\cite{hu2025_e2epref}.
In speech enhancement, URGENT-PK~\cite{wangURGENTPKPerceptuallyAlignedRanking2025} employs a pairwise model that directly compares homologous enhanced utterances.
To mitigate annotation scarcity, some approaches derive pairwise supervision from existing MOS datasets~\cite{shi2025_UPPSQA}. Preference-based modeling has also been explored under weak or indirect supervision, such as comparative assessment using non-matching reference signals~\cite{manocha2021noresqa}. In speech synthesis, pairwise preference prediction has been studied for samples sharing the same linguistic content to enforce consistency in relative judgments~\cite{valentini2022_predicting_pairwise,hu2023_preference_based_training}.

\subsection{Speech Quality Assessment with Natural Language Reasoning}

Recent work has begun to explore natural language reasoning for speech quality assessment, shifting from scalar score prediction toward language-based evaluation that better reflects human perceptual judgments. PAM prompts pretrained audio--language models with structured textual queries to assess speech quality, demonstrating strong zero-shot capability without task-specific training~\cite{deshmukh2024pam}. SpeechLLM-as-Judges further leverages large speech--language models as general-purpose evaluators, emphasizing interpretability through natural language explanations alongside quality judgments~\cite{SpeechJudge}. SpeechJudge advances this paradigm by training dedicated judgment models to achieve human-level performance in evaluating speech naturalness via language-based reasoning~\cite{zhang2025speechjudge}. Nevertheless, these methods remain largely experimental, with limited cross-domain generalization.

Complementing these model-centric approaches, QualiSpeech introduces a large-scale dataset with natural language descriptions and reasoning annotations, enabling systematic study of explainable speech quality assessment~\cite{wang2025qualispeech}.

\section{Method}

\begin{figure*}[t] 
    \centering
    \includegraphics[width=.9\linewidth]{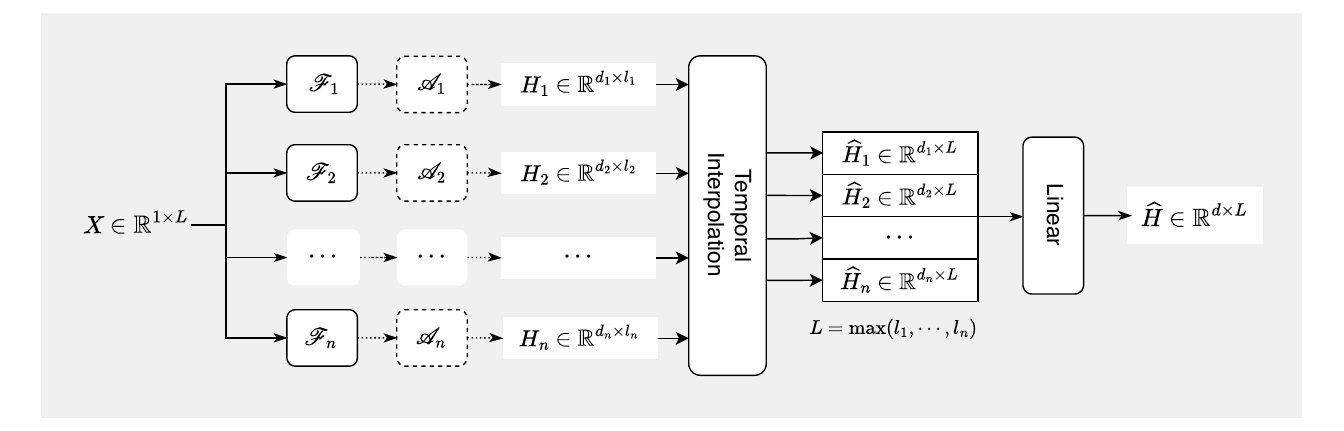} 
    \caption{Feature extractor in UrgentMOS. Representations  are temporally aligned via interpolation before fusion.}
    \label{im:feature_extractor}
    \vspace{-1em}
\end{figure*}

The overall architecture and training pipeline of UrgentMOS are illustrated in Figure~\ref{im:overall}. UrgentMOS adopts a modular design that jointly supports absolute and comparative speech quality assessment within a unified framework. Given a pair of speech samples \(X_A\) and \(X_B\), each signal is first processed independently by a shared feature extractor to obtain high-level acoustic representations, which are then used by both evaluation modules.

Absolute quality assessment is performed by the \textit{Absolute Metric Prediction Module (AMPM)}, which is shared across both inputs. 
AMPM follows a multi-metric prediction paradigm similar to Uni-VERSA~\cite{Uni-VERSA}, but differs in both architectural design and training formulation. 
Specifically, AMPM organizes speech quality metrics into semantically coherent groups \(C_1, \dots, C_m\) (Table~\ref{tab:metrics}), each corresponding to a distinct quality aspect and containing multiple related metrics. 
For each group \(C_i\), a \textit{shared metric encoder} transforms the acoustic representation into a group-specific latent representation \(\mathcal{H}^{A}_{C_i}\) or \(\mathcal{H}^{B}_{C_i}\) for inputs \(X_A\) and \(X_B\), respectively. 
Metric-specific prediction heads are applied to pooled representations \(\mathcal{H}_{C_i}\) to produce individual predictions 
, which are further constrained by range-constraining activation functions (rcAct) to ensure valid outputs while maintaining smooth optimization.


In addition to absolute metric prediction, UrgentMOS explicitly models relative quality judgments through a \textit{Naturalness-Conditioned Preference Module (NCPM)}. NCPM operates on representations from the naturalness-related metric group \(C_m\), which includes human MOS and learned naturalness predictors. Given the corresponding representations \(\mathcal{H}^{A}_{C_m}\) and \(\mathcal{H}^{B}_{C_m}\) from two speech samples, NCPM employs cross-attention to model interactions between paired inputs and produces comparative quality predictions, such as preference labels or comparative MOS (CMOS).

\subsection{Model Components}

\subsubsection{Feature Extractor}
\label{sec:feature_extractor}

UrgentMOS adopts a multi-branch, architecture-agnostic feature extractor to obtain complementary speech representations at different temporal resolutions, as shown in Figure~\ref{im:feature_extractor}. Given an input waveform \( X \), the signal is processed in parallel by a set of heterogeneous feature encoders \( \{ \mathcal{F}_i \}_{i=1}^{n} \). Each encoder produces a sequence of layer-wise representations, which are optionally aggregated by an aggregation module \( \mathcal{A}_i \) using a weighted sum. The resulting representation is denoted as \( H_i \in \mathbb{R}^{d_i \times l_i} \), where \( d_i \) and \( l_i \) correspond to the encoder-specific feature dimension and temporal length.

All representations are temporally aligned via length interpolation to a common sequence length \( L = \max_i l_i \), yielding aligned features \( \hat{H}_i \in \mathbb{R}^{d_i \times L} \). 
The aligned features are then fused into a unified representation \( \hat{H} \), which serves as the shared input to downstream absolute metric prediction and preference modeling modules. 
This design enables multi-resolution modeling and supports seamless integration of diverse pretrained audio encoders, including WavLM~\cite{WavLM-Chen2022}, Kimi-Audio~\cite{ding2025kimi}, Qwen3OmniCaptioner~\cite{qwen3omni-xu2025}, and Audio Flamingo~\cite{AF3-goel2025}.

\subsubsection{Range-Constraining Activations (rcAct)}

To ensure that predicted metric values lie within their valid numerical ranges, UrgentMOS applies a \textit{Range-Constraining Activation Module} after each metric-specific prediction head. Given an unconstrained scalar prediction \( x \), the module enforces lower and/or upper bounds according to the target metric’s domain. When both lower and upper bounds are finite, a scaled sigmoid function is used:
\begin{equation}
y = v_{\min} + (v_{\max} - v_{\min}) \cdot \sigma(x),
\end{equation}
which smoothly maps predictions to the interval \([v_{\min}, v_{\max}]\).
When only a lower bound is defined, the output is constrained via a softplus-based transformation:
\begin{equation}
y = v_{\min} + \mathrm{softplus}(x - v_{\min}),
\end{equation}
ensuring \( y \ge v_{\min} \).
Conversely, when only an upper bound is specified, the output is given by:
\begin{equation}
y = v_{\max} - \mathrm{softplus}(v_{\max} - x),
\end{equation}
which enforces \( y \le v_{\max} \).
If the output shall be unbounded, the identity mapping is applied. Compared to hard clipping, this formulation preserves gradient continuity near the boundaries, avoiding optimization instability while respecting metric-specific value constraints.

\begin{table*}[t]
\centering
\caption{Predicted Categorized Metrics in UrgentMOS. ``Ref.'' indicates whether the metric requires a reference signal for computation.}
\begin{tabular}{rl|l|c|c}
\hline
\textbf{\#} & \textbf{Category} & \textbf{Metric} & \textbf{Range} & \textbf{Ref.} \\
\hline
1  & \multirow{5}{*}{Noise \& Distortion} & PESQ~\cite{ITU-P862,PESQ-Rix2001} & [1, 4.5] & \checkmark \\
2  & & PESQc2~\cite{ITU-P862,PESQc2-Torcoli2025} & [1, 4.5] & \checkmark \\
3  & & DNSMOS~\cite{DNSMOS-Reddy2022} & [1, 5] & \xmark \\
4  & & LSD~\cite{LSD-Gray1976} & [0, $\infty$) & \checkmark \\
5  & & SDR~\cite{SDR-Vincent2006} & ($-\infty$, $\infty$) & \checkmark \\
\hline
6  & \multirow{5}{*}{Naturalness} & MOS (human) & [1, 5] & \xmark \\
7  & & UTMOS~\cite{UTMOS-Saeki2022} & [1, 5] & \xmark \\
8  & & Distill\_MOS~\cite{Distill_MOS-Stahl2025} & [1, 5] & \xmark \\
9  & & NISQA\_MOS~\cite{NISQA-Mittag2021} & [1, 5] & \xmark \\
10 & & SCOREQ~\cite{SCOREQ-Ragano2024} & [1, 5] & \xmark \\
\hline
11 & \multirow{3}{*}{Intelligibility} & ESTOI~\cite{ESTOI-Jensen2016} & [0, 1] & \checkmark \\
12 & & SpeechBERTScore~\cite{SpeechBERTScore-Saeki2024} & [-1, 1] & \checkmark \\
13 & & LPS~\cite{LPS-Pirklbauer2023} & ($-\infty$ ,1] & \checkmark \\
\hline
14 & Speaker Characteristics & SpeakerSimilarity & [--1, 1] & \checkmark \\
\hline
15 & Spectral Accuracy & MCD~\cite{MCD-Kubichek1993} & [0, $\infty$) & \checkmark \\
\hline
\end{tabular}
\label{tab:metrics}
\vspace{-1em}
\end{table*}

\subsection{Training Data Construction and Supervision}

\subsubsection{Learning with Incomplete Metric Annotations}

UrgentMOS leverages multi-metric supervision to improve robustness and generalization; however, in practice, different datasets provide heterogeneous subsets of quality annotations. Intrusive metrics require reference signals, while perceptual labels such as MOS are often unavailable at scale. Discarding samples with incomplete annotations would significantly reduce training data and domain coverage.

To address this, UrgentMOS explicitly tolerates missing metric annotations during training. For each metric \(k\), unavailable labels are marked as \texttt{NaN} and excluded via a binary validity mask \( m_k \in \{0,1\}^B \), where \(B\) is the batch size and \(m_k^{(b)}=1\) indicates that the ground-truth label \(y_k^{(b)}\) is available. The metric-specific loss is computed only over valid samples and normalized:
\begin{equation}
\label{eq:metric_loss}
\mathcal{L}_{MSE}^k
=
w_k
\frac{\sum_{b=1}^{B} m_k^{(b)} \,
\ell\!\left(\hat{y}_k^{(b)}, y_k^{(b)}\right)}
{\sum_{b=1}^{B} m_k^{(b)}},
\end{equation}
where \( \hat{y}_k^{(b)} \) denotes the prediction, \( \ell(\cdot,\cdot) \) is the regression loss, and \( w_k \) is a metric-specific weight. The overall objective averages losses over metrics with at least one valid label in the batch:
\begin{equation}
\mathcal{L}_{MSE}
=
\frac{1}{|\mathcal{K}_{\mathrm{valid}}|}
\sum_{k \in \mathcal{K}_{\mathrm{valid}}} \mathcal{L}_{MSE}^k,
\label{eq:loss_mse}
\end{equation}
where \( \mathcal{K}_{\mathrm{valid}} \) denotes the set of valid metrics. 

\subsubsection{Deriving Preference Supervision from ACR Annotations}
\label{sec:ccr_from_acr}

Comparative evaluation protocols such as CCR are widely adopted in speech quality assessment, yet publicly available CCR datasets remain limited in scale. In contrast, ACR datasets with MOS annotations are more abundant. UrgentMOS therefore derives preference supervision from existing ACR data by constructing sample pairs and assigning relative quality labels based on MOS differences.

Given two speech samples \(X_A\) and \(X_B\) with MOS scores \(s_A\) and \(s_B\), preference pairs are generated under different matching strategies. \emph{Arbitrary matching} (\(\mathcal{D}_{\text{any}}\)) allows pairing any two samples. To mitigate biases from dataset-specific annotation scales, \emph{corpus-level matching} (\(\mathcal{D}_{\text{corpus}}\)) restricts pairs to samples from the same dataset. \emph{Reference-level matching} (\(\mathcal{D}_{\text{ref}}\)) further constrains pairs to samples produced by different systems but sharing the same reference content.

To account for annotation noise and uncertainty in MOS values, a tunable tie threshold \( \delta > 0 \) is introduced. Preference labels are assigned as
\begin{equation}
y_{A,B} =
\begin{cases}
A \succ B, & s_A - s_B > \delta, \\
B \succ A, & s_B - s_A > \delta, \\
\text{tie}, & |s_A - s_B| \le \delta,
\end{cases}
\label{eq:ccr_from_acr}
\end{equation}
where \( y_{A,B} \) denotes the derived preference relation.

\subsubsection{Symmetric Pair Construction for Preference Learning}

Pairwise preference prediction should ideally be invariant to the ordering of input samples. While this property can be enforced through specialized network architectures, such constraints may unnecessarily limit model capacity and flexibility. UrgentMOS instead adopts a data-driven approach that encourages order-invariant behavior without modifying the model architecture. Concretely, for each preference-annotated pair $(X_A, X_B)$, we additionally construct its symmetric counterpart $(X_B, X_A)$ with a reversed preference label.
\footnote{Potential inconsistencies from independently annotated bidirectional preference pairs are not explicitly handled, as the datasets used either do not contain such paired annotations or do not provide sufficient metadata to reliably identify them.}

\begin{table*}[t]
\centering
\caption{Training datasets used in UrgentMOS. ``pairs'' refers to preference-labeled sample pairs.}
\begin{tabular}{l l S r r}
\hline
Dataset & Domain & \multicolumn{1}{l}{Samples} & Hours & Systems \\
\hline
BC19~\cite{BC19}                    & TTS                     & 136               & 0.32   & 21  \\
BVCC~\cite{BVCC}                    & VC                      & 4973           & 5.56   & 175 \\
NISQA~\cite{NISQA-Mittag2021}                  & Simulated               & 11020          & 27.21  & --  \\
PSTN~\cite{PSTN}                    & Simulated               & 58709          & 163.08 & --  \\
SOMOS~\cite{SOMOS}                  & TTS                     & 14100          & 18.32  & 181 \\
TCD-VoIP~\cite{TCD-VoIP}            & Simulated (VoIP)        & 384               & 0.87   & 24  \\
Tencent~\cite{Tencent}              & SE Enhanced             & 11563          & 23.51  & --  \\
TMHINT-QI~\cite{TMHINT-QI}           & Simulated (Noise)       & 12937          & 11.35  & 98  \\
TTSDS2~\cite{TTSDS2}                & TTS                     & 460               & 0.96   & 80  \\
URGENT2024-SQA~\cite{URGENT-SQA}    & SE Enhanced             & 238000         & 429.34 & 238 \\
URGENT2025-SQA~\cite{URGENT-SQA}    & SE Enhanced             & 100000         & 261.31 & 100 \\
SpeechEval~\cite{SpeechEval}        & Mixed                   & \multicolumn{1}{c}{73,123 pairs}     & 18.25  & 15  \\
SpeechJudge-Data~\cite{SpeechJudge} & Mixed                   & \multicolumn{1}{c}{41,755 pairs}    & 328.50 & --  \\
\hline
\end{tabular}
\label{tab:training_data}
\vspace{-1.5em}
\end{table*}

\section{Experiment Setup}
\subsection{Datasets}

\subsubsection{Training}

UrgentMOS is trained on a diverse collection of speech quality datasets spanning text-to-speech (TTS), voice conversion (VC), speech enhancement (SE), and simulated distortion domains, as summarized in Table~\ref{tab:training_data}. The training corpus comprises both absolute-score annotated samples and preference-labeled pairs, enabling unified learning of absolute quality estimation and comparative assessment.  Absolute-score supervision covers synthesized speech (e.g., BC19, SOMOS, TTSDS2), voice conversion outputs (BVCC), simulated and real-world degradations such as telephony artifacts (PSTN, TCD-VoIP) and background noise (TMHINT-QI), as well as enhanced speech produced by SE systems (Tencent, URGENT2024-SQA, URGENT2025-SQA). 
In addition, preference-annotated datasets, including SpeechEval and SpeechJudge-Data, are incorporated to provide explicit comparative supervision.

\subsubsection{Evaluation}

Evaluation is conducted using a combination of officially released test sets and constructed preference pairs to reflect both absolute and comparative assessment scenarios. For datasets that provide standard evaluation splits, we directly adopt the official test sets.
To further align evaluation with practical preference-based assessment, we additionally construct pairwise comparison data using \emph{reference-scope matching}, where samples generated under the same reference content are paired to form comparative judgments. This construction is applied only to datasets that contain sufficient system and reference information to support such pairing, as summarized in Table~\ref{tab:evaluation_data}. For large-scale datasets, the total number of constructed preference pairs is capped at 100{,}000.

\begin{table}[t]
\centering
\caption{Constructed preference pairs using reference-scope matching for evaluation.}
\begin{tabular}{l S}
\hline
\textbf{Dataset} & \multicolumn{1}{l}{\textbf{\#Preference Pairs}} \\
\hline
NISQA P501                         & 14160 \\
NISQA FOR                     & 6496  \\
SOMOS                          & 7250  \\
TMHINT-QI                            & 16708 \\
CHiME-7-UDASE-Eval                   & 2560  \\
URGENT2024-SQA     & 100000 \\
URGENT2025-SQA     & 100000 \\
\hline
\end{tabular}
\label{tab:evaluation_data}
\vspace{-1.5em}
\end{table}

\subsection{Metrics}

We evaluate UrgentMOS using preference-based and absolute-score metrics. For pairwise evaluation, we report preference prediction accuracy over \(\{A \succ B, B \succ A, \text{tie}\}\). For absolute evaluation, we report utterance-level Pearson correlation (LCC) and Spearman rank correlation (SRCC) between predicted scores and ground-truth MOS.

\subsection{Training Details}
Each shared metric encoder in UrgentMOS is implemented as a Transformer with 6 layers, 8 attention heads, a model dimension of 768, and a feed-forward dimension of 2048. The weights of the multiple metrics in the loss computation in (\ref{eq:metric_loss}) are set to be equal. The model is trained using a constant learning rate of \(3\times10^{-5}\). We adopt a dynamic batching strategy based on audio duration, with each batch containing approximately 400 seconds of speech. Training is conducted for 20k optimization steps on a single NVIDIA A100 GPU. Depending on the number of feature extractors, the total training time ranges from 3 to 20 hours. 
\section{Results}

UrgentMOS variants in Table~\ref{tab:eval_pair_results} are denoted as \textbf{F\#C\#M\#--$\mathcal{D}_{*}$}. 
\textbf{F\#} indicates the number of feature extractors (F1: WavLM only; F4: four extractors described in Section~\ref{sec:feature_extractor}); 
\textbf{C\#} denotes the number of metric categories (C1: all metrics treated as one group; C5: categorized as in Table~\ref{tab:metrics}); 
\textbf{M\#} specifies the supervision scope (M1: MOS only; M5: naturalness-related metrics; M15: all metrics listed in Table~\ref{tab:metrics}). 
$\mathcal{D}_{*}$ denotes the preference-pair construction strategy in \ref{sec:ccr_from_acr}. Training and inference statistics are available in Appendix ~\ref{appendix:computational_cost}.

\begin{table*}[t]
\centering
\caption{Preference evaluation accuracy (acc$_{0.5}$ / acc$_0$). Best results are shown in \textbf{bold} and second-best results are \underline{underlined}. Additional results are reported in Table~\ref{tab:eval_pair_results_appendix} in Appendix~\ref{appendix:additional_evaluation_results}.}
\setlength{\tabcolsep}{6pt}
\renewcommand{\arraystretch}{1.1}
\begin{tabular}{l c c c c c}
\hline
\textbf{Model} 
& \textbf{SOMOS} 
& \textbf{TMHINT-QI} 
& \textbf{URGENT24-SQA} 
& \textbf{SpeechEval} 
& \textbf{SpeechJudge} \\
\hline
DNSMOS       
& 0.49 / 0.51 
& 0.43 / 0.62 
& 0.54 / 0.63 
& 0.52 / 0.73 
& 0.07 / 0.58 \\

UTMOS         
& 0.52 / 0.64 
& 0.46 / 0.68 
& 0.55 / 0.67 
& 0.68 / 0.76 
& 0.23 / 0.60 \\

SCOREQ        
& 0.51 / 0.61 
& 0.48 / 0.63 
& 0.58 / 0.68 
& 0.65 / 0.76 
& 0.12 / 0.58 \\

Distill-MOS   
& 0.48 / 0.51 
& 0.47 / 0.63 
& 0.54 / 0.66 
& 0.64 / 0.74 
& 0.07 / 0.58 \\

NISQA-MOS     
& 0.47 / 0.51 
& 0.46 / 0.60 
& 0.55 / 0.67 
& 0.58 / 0.71 
& 0.18 / 0.58 \\

SpeechEval    
& 0.35 / 0.68
& 0.57 / \textbf{0.84} 
& 0.43 / 0.67
& 0.79 / 0.87
& 0.45 / 0.55 \\

SpeechJudge   
& 0.28 / 0.56 
& 0.50 / 0.70 
& 0.29 / 0.54 
& 0.48 / 0.59 
& \underline{0.74 / 0.74} \\

Random        
& 0.34 / 0.50 
& 0.34 / 0.50 
& 0.33 / 0.50 
& 0.33 / 0.50 
& 0.30 / 0.52 \\

\hline
\multicolumn{6}{c}{\textbf{UrgentMOS}} \\
\hline
F1C1M1-$\mathcal{D}_{\mathrm{ref}}$     
& 0.57 / 0.71
& 0.60 / 0.77 
& 0.57 / 0.67 
& 0.80 / 0.87
& 0.67 / 0.67 \\

F1C1M5-$\mathcal{D}_{\mathrm{ref}}$     
& 0.58 / 0.72 
& \textbf{0.63} / 0.77 
& 0.58 / 0.69 
& 0.81 / 0.89 
& 0.68 / 0.68 \\

F1C1M5-$\mathcal{D}_{\mathrm{corpus}}$     
& \underline{0.59 / 0.72 }
& 0.62 / 0.77 
& 0.58 / 0.68 
& 0.81 / 0.89 
& 0.63 / 0.63 \\

F1C1M5-$\mathcal{D}_{\mathrm{any}}$     
& 0.58 / 0.71 
& 0.62 / 0.77 
& \textbf{0.59 / 0.69}
& 0.81 / \textbf{0.91}
& 0.69 / 0.69 \\

F4C1M5-$\mathcal{D}_{\mathrm{ref}}$
& \textbf{0.60 / 0.73} 
& \textbf{0.63} / \underline{0.78} 
& \textbf{0.59 / 0.69}
& \textbf{0.83 / 0.91} 
& \textbf{0.75 / 0.75} \\

F1C1M15-$\mathcal{D}_{\mathrm{ref}}$     
& 0.57 / 0.69 
& 0.58 / 0.75 
& 0.57 / 0.68 
& 0.78 / 0.90 
& 0.65 / 0.65 \\

F1C5M15-$\mathcal{D}_{\mathrm{ref}}$     
& 0.57 / 0.70 
& 0.62 / 0.76 
& \textbf{0.59 / 0.69} 
& \underline{0.82} / 0.89 
& 0.71 / 0.71 \\

\hline
\end{tabular}
\label{tab:eval_pair_results}
\vspace{-.5em}
\end{table*}

\begin{table*}[t!]
\centering
\caption{Correlation results (LCC / SRCC) on representative datasets. Best results are shown in \textbf{bold} and second-best results are \underline{underlined}. Additional results are reported in Table~\ref{tab:eval_corr_results_appendix} in Appendix~\ref{appendix:additional_evaluation_results}.}
\setlength{\tabcolsep}{6pt}
\begin{tabular}{l c c c c c}
\hline
\textbf{Model} 
& \textbf{SOMOS} 
& \textbf{TMHINT-QI} 
& \textbf{URGENT24-SQA} 
& \textbf{BVCC} 
& \textbf{NISQA-FOR} \\
\hline
DNSMOS       
& 0.01 / 0.01 
& 0.41 / 0.37 
& 0.65 / 0.64 
& 0.49 / 0.50 
& 0.54 / 0.54 \\

UTMOS         
& 0.44 / 0.42 
& 0.60 / 0.48 
& 0.72 / 0.74 
& \textbf{0.88 / 0.88} 
& 0.80 / 0.77 \\

SCOREQ        
& 0.32 / 0.31 
& 0.62 / 0.47 
& \underline{0.79 / 0.79} 
& 0.81 / 0.82 
& \underline{0.93 / 0.93} \\

Distill-MOS   
& 0.00 / 0.01 
& 0.57 / 0.46 
& 0.76 / 0.75 
& 0.82 / 0.81 
& 0.82 / 0.83 \\

NISQA-MOS     
& 0.02 / 0.03 
& 0.53 / 0.35 
& 0.71 / 0.70 
& 0.62 / 0.63 
& 0.89 / 0.88 \\

\hline
\multicolumn{6}{c}{\textbf{UrgentMOS}} \\
\hline
F1C1M1-$\mathcal{D}_{\mathrm{ref}}$     
& 0.64 / 0.63 
& 0.79 / 0.74 
& 0.75 / 0.75 
& 0.83 / 0.83 
& 0.91 / 0.90 \\

F1C1M5-$\mathcal{D}_{\mathrm{ref}}$     
& \textbf{0.68 / 0.67} 
& 0.80 / 0.76 
& 0.78 / 0.78
& 0.84 / 0.85 
& 0.92 / 0.91 \\

F1C1M5-$\mathcal{D}_{\mathrm{corpus}}$     
& 0.65 / 0.63 
& 0.80 / 0.76 
& 0.77 / 0.77 
& \underline{0.86 / 0.86} 
& 0.93 / 0.92 \\

F1C1M5-$\mathcal{D}_{\mathrm{any}}$     
& 0.65 / 0.64
& 0.80 / 0.76 
& 0.78 / 0.77 
& 0.85 / 0.85 
& 0.91 / 0.90 \\

F4C1M5-$\mathcal{D}_{\mathrm{ref}}$     
& \textbf{0.68 / 0.67} 
& \textbf{0.82 / 0.78} 
& \textbf{0.80 / 0.79} 
& \underline{0.86 / 0.86}
& \textbf{0.93 / 0.94} \\

F1C1M15-$\mathcal{D}_{\mathrm{ref}}$     
& 0.57 / 0.55 
& 0.78 / 0.72 
& 0.76 / 0.76 
& 0.82 / 0.82 
& 0.90 / 0.90 \\

F1C5M15-$\mathcal{D}_{\mathrm{ref}}$     
& 0.62 / 0.61
& 0.80 / 0.75
& 0.79 / 0.78
& 0.85 / 0.85
& 0.93 / 0.92 \\

\hline
\end{tabular}
\label{tab:eval_corr_results}
\vspace{-1.5em}
\end{table*}

\subsection{Preference-based Evaluation}

Table~\ref{tab:eval_pair_results} reports preference prediction accuracy under two settings, acc$_{0.5}$ and acc$_0$. For natively preference-based datasets (SpeechEval and SpeechJudge), acc$_{0.5}$ allows tie predictions, whereas acc$_0$ evaluates only strictly ordered pairs by removing ties. For derived preference datasets, acc$_{0.5}$ uses a threshold $\delta = 0.5$ to treat pairs with small score differences as ties, while acc$_0$ sets $\delta = 0$ and evaluates all pairs as strict preferences (see (\ref{eq:ccr_from_acr})). 

As shown in Table~\ref{tab:eval_pair_results}, SpeechEval and SpeechJudge models perform well on their own benchmarks but show noticeably weaker generalization to other domains., indicating dataset-specific overfitting. In contrast, UrgentMOS variants consistently achieve strong performance across TTS, simulated noise, speech enhancement, and mixed-domain datasets, demonstrating better robustness.

Notably, most existing MOS predictors perform worse than random under acc$_{0.5}$ on the SpeechJudge test set, which consists of synthesized speech from recent TTS systems. This behavior can be attributed to score compression: for modern, high-quality samples, these predictors tend to output highly similar scores, leading to a large proportion of predicted ties when $\delta = 0.5$ is applied. Because the SpeechJudge test set does not include tie annotations, such ambiguous predictions are penalized, indicating that existing MOS models have limited sensitivity to the subtle quality differences exhibited by state-of-the-art TTS systems.

Interestingly, the choice of preference construction strategy has limited impact on UrgentMOS performance. Models trained using reference-level ($\mathcal{D}_\mathrm{ref}$), corpus-level ($\mathcal{D}_\mathrm{corpus}$), or unconstrained ($\mathcal{D}_\mathrm{any}$) pairing achieve comparable accuracy across evaluation datasets. This suggests that a threshold of $\sigma = 0.5$ during training is sufficient to separate perceptually different samples, even when cross-corpus inconsistencies exist.

Moreover, models trained with naturalness-related multi-metric supervision (C5) consistently outperform their MOS-only counterparts (C1), indicating that incorporating complementary perceptual cues substantially improves preference discrimination. In contrast, extending supervision to all 15 metrics does not always yield additional gains. Notably, F1C1M15 shows degraded performance on several datasets, particularly when a single shared feature encoder is used across all metrics. This degradation is likely due to weak or inconsistent correlations among certain metrics, as evidenced by the correlation analysis in Appendix~\ref{appendix:corr_heatmap}. Nevertheless, the broader multi-metric prediction capability of the M15 variants remains valuable, as it provides rich, multi-aspect quality signals that may support more informative audio quality reasoning in LLM-based speech assessment frameworks.

\subsection{Correlation-based Evaluation}

Table~\ref{tab:eval_corr_results} presents LCC and SRCC on datasets with absolute MOS annotations; 
The correlation results largely align with the preference accuracy trends.

UrgentMOS consistently achieves strong correlations across all datasets, demonstrating better cross-domain robustness than conventional MOS predictors. 
Naturalness-focused multi-metric supervision (M5) and multi-feature extraction (F4) provide consistent gains, while incorporating all metrics (M15) does not always yield further improvements, likely due to weaker inter-metric correlations. 

\paragraph{Ablation Study and Analysis}

Table~\ref{tab:encoder_eval_corr_results_appendix} evaluates encoders adopted in F4C1M5-$\mathcal{D}_{\mathrm{ref}}$. Ablation studies on symmetric pair construction are reported in Appendix~\ref{appendix:ablation_symmetric_pairs}. Analysis on the CMOS threshold for preference prediction with post-hoc MOS scores is provided in Appendix~\ref{appendix:preference_threshold_analysis}.
Analysis on other predicted metrics are provided in Appendix~\ref{appendix:additional_metric_results}.

\vspace{-.5em}
\section{Conclusion}
\vspace{-.5em}

In this work, we presented UrgentMOS, a unified speech quality assessment framework designed to address the limitations of existing MOS-based evaluators in modern speech generation scenarios. UrgentMOS learns from heterogeneous objective and perceptual metrics while explicitly tolerating missing annotations, enabling robust training on large-scale, multi-source datasets with inconsistent supervision. By integrating absolute metric prediction with naturalness-conditioned preference modeling, it supports both utterance-level quality estimation and pairwise preference evaluation, aligning model training with practical benchmarking protocols. Experiments across diverse domains show that UrgentMOS consistently outperforms strong baselines.  


\section*{Limitations}

Despite its strong performance, UrgentMOS does not provide explicit natural language explanations for its quality judgments, which contrasts with recent language-based evaluation approaches. Nevertheless, its ability to predict multiple complementary quality metrics offers structured cues that may support downstream reasoning or explainability modules.

Incorporating multiple feature extractors improves robustness but increases inference cost, potentially limiting the use of UrgentMOS in large-scale or latency-sensitive data filtering scenarios. In addition, while multi-metric supervision generally enhances generalization, introducing a large number of heterogeneous metrics can degrade performance in some settings, particularly when correlations among metrics are weak or inconsistent. Identifying optimal metric subsets or developing training strategies to better resolve such conflicts remains an open direction.

\bibliography{custom}

\appendix

\newpage
\section{Correlation between Metrics}
\label{appendix:corr_heatmap}

\begin{figure}[h!] 
    \centering
    \includegraphics[width=\linewidth]{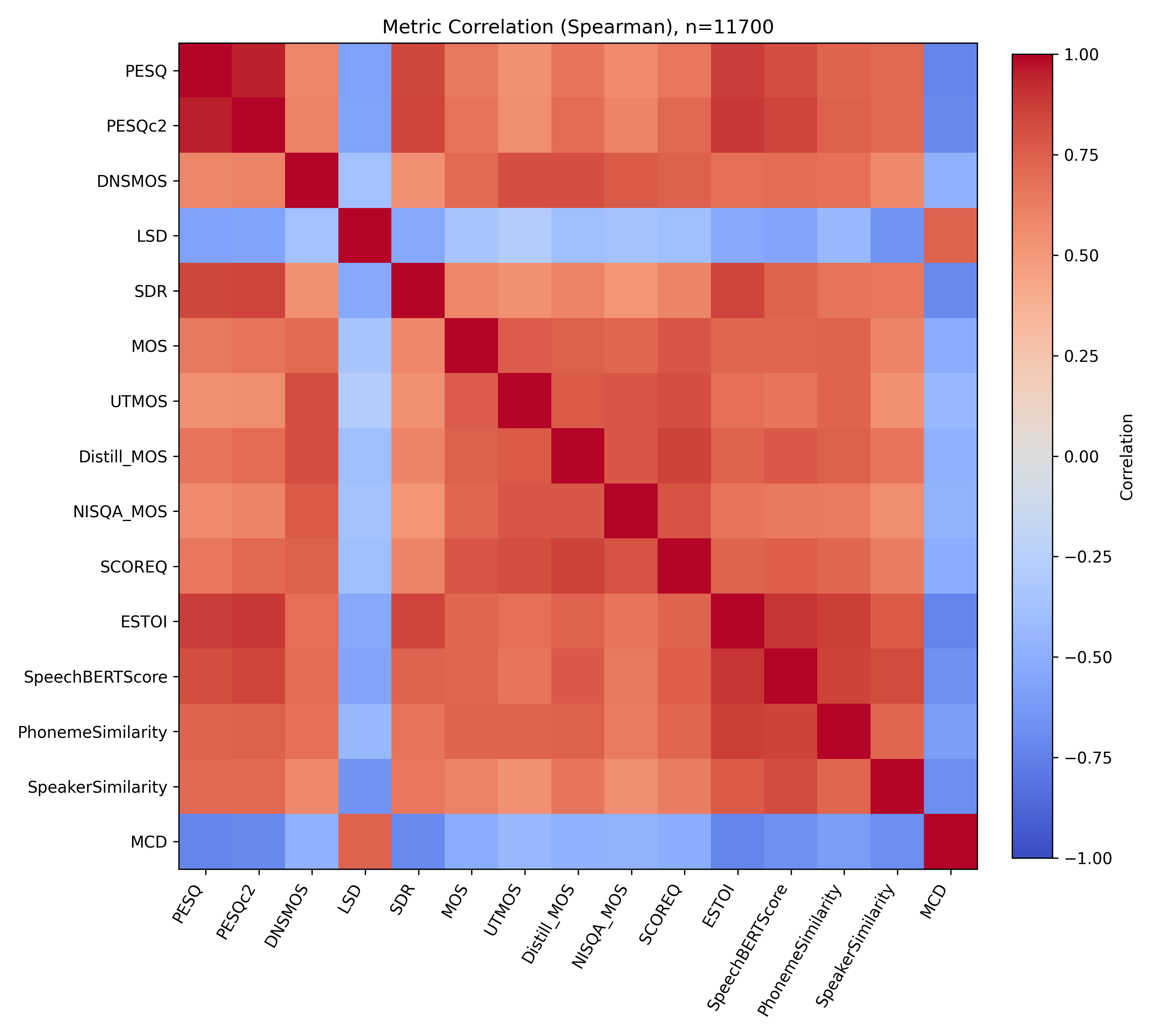}
    \caption{Spearman correlation across different speech quality metrics on the Urgent2025-SQA dataset.}
    \label{im:corr_heatmap}
\end{figure}

Figure~\ref{im:corr_heatmap} illustrates the correlation structure among different speech quality metrics. Most perceptual and learning-based metrics (e.g., MOS, UTMOS, Distill-MOS, NISQA-MOS, ESTOI, and semantic or speaker similarity scores) form a strongly correlated cluster, suggesting that they capture largely overlapping perceptual cues. In contrast, several signal-level distortion metrics, notably LSD and MCD, show much weaker or even negative correlations with MOS and related perceptual measures.

This disparity partially explains why predicting a larger set of metrics can degrade performance. Jointly optimizing metrics that are weakly aligned with human perception introduces conflicting supervision, which can interfere with learning a coherent quality representation. Consequently, simply increasing the number of predicted metrics does not guarantee gains.

\section{Computational Cost}
\label{appendix:computational_cost}

\subsection{Training Efficiency}

Table~\ref{tab:training_efficiency} reports the training efficiency of UrgentMOS measured in iterations per second. 
Configurations using a single feature extractor (F1) achieve stable training throughput, with only minor overhead introduced by additional metric groups. 
Training speed decreases noticeably when using larger or computationally heavier encoders (e.g., Kimi-Audio, AF-Whisper), and drops most significantly in the multi-encoder setting (F4C1M5) due to parallel feature extraction. 

\begin{table}[h!]
\centering
\caption{Training Efficiency of UrgentMOS}
\setlength{\tabcolsep}{8pt}
\begin{tabular}{l c}
\hline
\textbf{Setting} & \textbf{iter / s} \\
\hline
F1C1M1 (WavLM).  & 2.30   \\
F1C1M5 (WavLM).  & 2.16    \\
F1C1M5 (Kimi-Audio).  & 0.74    \\
F1C1M5 (Qwen3).  & 2.02    \\
F1C1M5 (AF-Whisper).  & 0.68    \\
F1C1M15 (WavLM) &  2.12  \\
F1C5M15 (WavLM) &  1.38 \\
F4C1M5 &   0.32  \\
\hline
\end{tabular}
\vspace{-1em}
\label{tab:training_efficiency}
\end{table}

\subsection{Inference Efficiency}

Table~\ref{tab:inference_efficiency} summarizes inference efficiency in terms of real-time factor (RTF), defined as the ratio between processing time and audio duration, where lower values indicate faster inference. 
UrgentMOS achieves very low RTFs with a single feature extractor (\(\leq 0.003\)), while configurations with multiple extractors incur higher computational cost, as illustrated by the increased RTF of the F4C1M5 setting. 
Baseline results are reported using their respective open-source implementations and \textbf{do not necessarily reflect intrinsic model efficiency.} 
By contrast, UrgentMOS adopts inference-oriented optimizations such as padding-efficient batching, and the reported numbers reflect practical deployment efficiency under direct, off-the-shelf usage.

\begin{table}[h!]
\centering
\caption{Inference Efficiency of UrgentMOS}
\begin{tabular}{l c}
\hline
\textbf{Setting} & \textbf{real time factor (RTF)} \\
\hline
DNSMOS  & 0.011    \\
UTMOS  &  0.006   \\
SCOREQ  &   0.027  \\
Distill-MOS  &  0.005   \\
NISQA-MOS  &  0.031   \\
SpeechEval  &  0.018  \\
SpeechJudge  &  0.037   \\
\hline
\multicolumn{2}{c}{UrgentMOS} \\
\hline
F1C1M1 (WavLM)  &  0.001   \\
F1C1M5 (WavLM)  &  0.002   \\
F1C1M5 (Kimi-Audio)  &  0.002   \\
F1C1M5 (Qwen3)  &  0.002   \\
F1C1M5 (AF-Whisper)  &  0.003   \\
F1C1M15 (WavLM) &  0.002  \\
F1C5M15 (WavLM) & 0.002  \\
F4C1M5 &   0.009  \\
\hline
\end{tabular}
\vspace{-1em}
\label{tab:inference_efficiency}
\end{table}

\section{Additional Metric Correlation Results}
\label{appendix:additional_metric_results}

Table~\ref{tab:metric_corr_nonmos} reports correlations between predicted and ground-truth quality metrics on the URGENT-SQA test sets, excluding MOS. UrgentMOS achieves consistently strong correlations across diverse metric families, including distortion-related (e.g., PESQ, SDR), spectral (e.g., MCD), intelligibility (e.g., ESTOI, SpeechBERTScore), and speaker similarity metrics, indicating that it can reliably recover multiple complementary aspects of speech quality within a unified framework.  

Although the MOS correlation of the full multi-metric configuration (F1C5M15-$\mathcal{D}_{ref}$) is not always higher than that of F1C1M5-$\mathcal{D}_{ref}$, the ability to jointly predict heterogeneous quality metrics remains valuable. These multi-aspect predictions provide structured and interpretable cues that extend beyond a single scalar score and can serve as informative signals for downstream analysis and natural language-based quality reasoning, where explicit access to different perceptual dimensions is beneficial.

\section{Ablation Study on Symmetric Pair Construction}
\label{appendix:ablation_symmetric_pairs}

\begin{table}[h!]
\centering
\caption{Effect of symmetric pair construction of F1C1M5-$\mathcal{D}_{\mathrm{ref}}$ on preference accuracy (acc$_{0.5}$ / acc$_0$).}
\begin{tabular}{l c c}
\hline
\textbf{Setting} & \textbf{SpeechEval} & \textbf{SpeechJudge} \\
\hline
w/ symm.  & 0.81 / 0.89   & 0.68 / 0.68 \\
w/o symm. & 0.79 / 0.85   & 0.63 / 0.63 \\
\hline
\end{tabular}
\vspace{-1em}
\label{tab:ablation_symmetric}
\end{table}

\section{Preference Pair Construction Threshold Analysis}
\label{appendix:preference_threshold_analysis}

\begin{figure*}[h] 
    \centering
    \includegraphics[width=.9\linewidth]{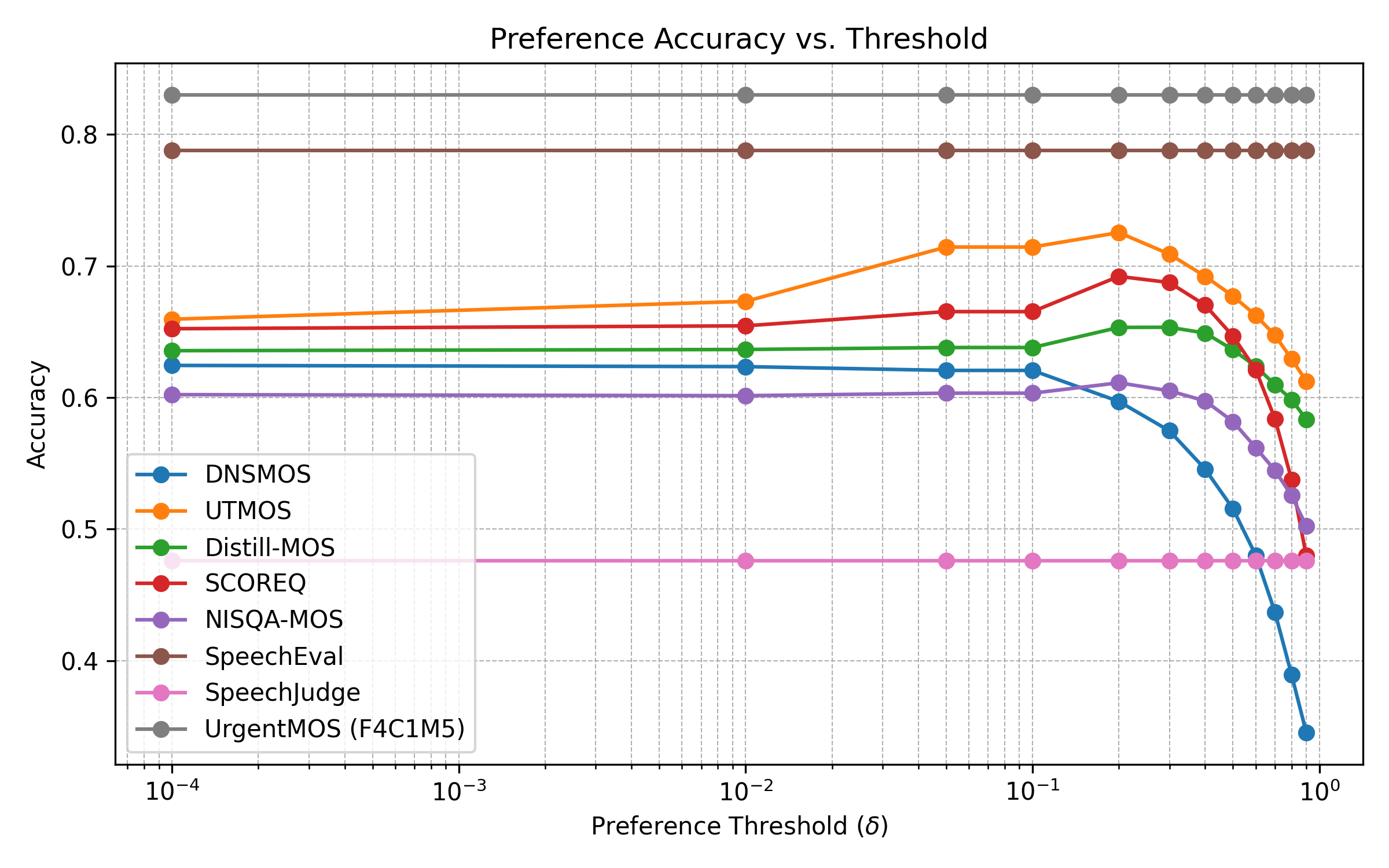} 
    \caption{Preference accuracy on the SpeechEval test set versus tie threshold~\(\delta\).
Direct preference models are threshold-invariant, while MOS predictors are highly sensitive.}
    \label{im:preference_accuracy_vs_threshold}
\end{figure*}

Figure~\ref{im:preference_accuracy_vs_threshold} illustrates preference accuracy as a function of the tie threshold~\(\delta\).
Models that directly predict pairwise preferences, including \textbf{UrgentMOS},
\textbf{SpeechEval}, and \textbf{SpeechJudge}, exhibit constant performance across
all thresholds, as they do not rely on score differences and are therefore
insensitive to \(\delta\).

In contrast, conventional MOS predictors are highly sensitive to the threshold
choice, with the optimal \(\delta\) varying across models. Even when an oracle
threshold is used to maximize their performance, a substantial gap remains
compared to UrgentMOS, highlighting the benefit of explicitly modeling
preferences rather than deriving them from absolute MOS scores.

\begin{table*}[h]
\centering
\caption{Preference evaluation accuracy (acc$_{0.5}$ / acc$_0$) on additional test sets not included in the main paper. \textbf{CHiME-7.} refers to the CHiME-7-UDASE-Eval set.}
\setlength{\tabcolsep}{6pt}
\renewcommand{\arraystretch}{1.1}
\begin{tabular}{l c c c c}
\hline
\textbf{Model} 
& \textbf{NISQA-P501} 
& \textbf{NISQA-FOR} 
& \textbf{CHiME-7.} 
& \textbf{URGENT25-SQA} \\
\hline
DNSMOS       
& 0.46 / 0.78 
& 0.47 / 0.69 
& 0.37 / 0.51 
& 0.52 / 0.63 \\

UTMOS         
& 0.72 / 0.85 
& 0.67 / 0.79 
& 0.49 / 0.57 
& 0.50 / 0.66 \\

SCOREQ        
& \textbf{0.81} / 0.90
& \underline{0.79 / 0.88}
& 0.60 / 0.75
& 0.57 / 0.69 \\

Distill-MOS   
& 0.75 / 0.86 
& 0.69 / 0.82 
& 0.55 / 0.67 
& 0.55 / 0.67 \\

NISQA-MOS     
& 0.74 / 0.87 
& 0.74 / 0.85 
& 0.51 / 0.59 
& 0.52 / 0.65 \\

SpeechEval    
& 0.68 / \textbf{0.91} 
& 0.53 / 0.77 
& 0.46 / 0.74 
& 0.50 / \textbf{0.78} \\

SpeechJudge   
& 0.54 / 0.71 
& 0.42 / 0.65 
& 0.30 / 0.58 
& 0.33 / 0.57 \\

Random        
& 0.33 / 0.50 
& 0.33 / 0.50 
& 0.34 / 0.51 
& 0.33 / 0.50 \\

\hline
\multicolumn{5}{c}{\textbf{UrgentMOS}} \\
\hline
F1C1M1-$\mathcal{D}_{\mathrm{ref}}$     
& 0.78 / 0.90 
& 0.76 / 0.86 
& 0.56 / 0.69 
& 0.57 / 0.69 \\

F1C1M5-$\mathcal{D}_{\mathrm{ref}}$     
& 0.79 / 0.89 
& 0.77 / 0.87 
& \underline{0.64} / 0.76 
& 0.59 / 0.70 \\

F1C1M5-$\mathcal{D}_{\mathrm{corpus}}$     
& 0.79 / 0.90 
& \underline{0.79 / 0.88} 
& \underline{0.64} / 0.77 
& 0.58 / 0.70 \\

F1C1M5-$\mathcal{D}_{\mathrm{any}}$     
& 0.80 / 0.90
& 0.76 / 0.86 
& \textbf{0.66} / \underline{0.78} 
& 0.58 / 0.70 \\

F4C1M5-$\mathcal{D}_{\mathrm{ref}}$     
& \textbf{0.81 / 0.91} 
& \textbf{0.81 / 0.89} 
& 0.60 / \textbf{0.80} 
& \textbf{0.59} / \underline{0.74} \\

F1C1M15-$\mathcal{D}_{\mathrm{ref}}$     
& 0.76 / 0.88 
& 0.75 / 0.86 
& 0.60 / 0.76 
& 0.58 / 0.69 \\

F1C5M15-$\mathcal{D}_{\mathrm{ref}}$     
& 0.80 / 0.90 
& \underline{0.79 / 0.88} 
& 0.62 / 0.77 
& \textbf{0.59} / 0.70 \\

\hline
\end{tabular}
\label{tab:eval_pair_results_appendix}
\end{table*}

\begin{table*}[h]
\centering
\caption{Additional correlation results (LCC / SRCC) on evaluation datasets not shown in the main paper. Best results are in \textbf{bold}, second-best are \underline{underlined}. \textbf{CHiME-7.} refers to the CHiME-7-UDASE-Eval set.}
\setlength{\tabcolsep}{6pt}
\renewcommand{\arraystretch}{1.1}
\begin{tabular}{l c c c c c}
\hline
\textbf{Model} 
& \textbf{BC19} 
& \textbf{NISQA-LIVE} 
& \textbf{NISQA-P501} 
& \textbf{CHiME-7.} 
& \textbf{URGENT25-SQA} \\
\hline
DNSMOS       
& 0.48 / 0.47 
& 0.60 / 0.59 
& 0.65 / 0.70 
& 0.28 / 0.28 
& 0.65 / 0.65 \\

UTMOS         
& 0.61 / 0.65 
& 0.77 / 0.77 
& 0.85 / 0.88 
& 0.26 / 0.33 
& 0.65 / 0.69 \\

SCOREQ        
& 0.69 / 0.66 
& 0.76 / 0.74 
& \textbf{0.93 / 0.94} 
& 0.75 / 0.75
& 0.74 / 0.74 \\

Distill-MOS   
& \underline{0.84 / 0.81} 
& 0.86 / 0.84
& 0.90 / 0.90 
& 0.72 / 0.71
& 0.71 / 0.70 \\

NISQA-MOS     
& 0.43 / 0.33 
& 0.78 / 0.76 
& 0.90 / 0.91
& 0.48 / 0.47 
& 0.68 / 0.66 \\

\hline
\multicolumn{6}{c}{\textbf{UrgentMOS}} \\
\hline
F1C1M1-$\mathcal{D}_{ref}$     
& 0.81 / 0.77 
& 0.83 / 0.81 
& 0.92 / 0.93 
& 0.64 / 0.63 
& 0.74 / 0.74 \\

F1C1M5-$\mathcal{D}_{ref}$     
& \textbf{0.87 / 0.84} 
& 0.83 / 0.82 
& 0.93 / 0.93
& 0.76 / 0.75 
& \textbf{0.76 / 0.76} \\

F1C1M5-$\mathcal{D}_{corpus}$     
& 0.80 / 0.77 
& 0.82 / 0.79 
& \textbf{0.93 / 0.94} 
& 0.76 / 0.75 
& 0.74 / 0.74 \\

F1C1M5-$\mathcal{D}_{any}$     
& 0.81 / 0.76 
& 0.83 / 0.80 
& 0.92 / 0.94 
& \textbf{0.79 / 0.78} 
& 0.74 / 0.73 \\

F4C1M5-$\mathcal{D}_{ref}$     
& 0.83 / 0.77 
& \textbf{0.87 / 0.87} 
& \textbf{0.93 / 0.94} 
& \underline{0.77 / 0.76} 
& 0.75 / 0.74 \\

F1C1M15-$\mathcal{D}_{ref}$     
& 0.74 / 0.69 
& \underline{0.87 / 0.85} 
& 0.90 / 0.92 
& 0.68 / 0.67 
& 0.75 / 0.74 \\

F1C5M15-$\mathcal{D}_{ref}$     
& 0.82 / 0.78
& 0.86 / 0.84
& 0.93 / 0.94
& 0.76 / 0.75
& \underline{0.76 / 0.75} \\

\hline
\end{tabular}
\label{tab:eval_corr_results_appendix}
\end{table*}
 
\begin{table*}[h!]
\centering
\caption{Correlation results (LCC / SRCC) for F1C1M5-$\mathcal{D}_{\mathrm{ref}}$ of UrgentMOS with different encoders on main and additional evaluation datasets.}
\setlength{\tabcolsep}{6pt}
\renewcommand{\arraystretch}{1.1}
\begin{tabular}{l c c c c c}
\hline
\textbf{Encoder} 
& \textbf{SOMOS} 
& \textbf{TMHINT-QI} 
& \textbf{URGENT24-SQA} 
& \textbf{BVCC} 
& \textbf{NISQA-FOR} \\
\hline
WavLM
& 0.64 / 0.63
& 0.80 / 0.76
& 0.78 / 0.78
& 0.84 / 0.85
& 0.92 / 0.91 \\

Kimi-Audio
& 0.68 / 0.67
& 0.81 / 0.77
& 0.77 / 0.78
& 0.85 / 0.85
& 0.93 / 0.93 \\

Qwen3Omni
& 0.60 / 0.59
& 0.78 / 0.74
& 0.78 / 0.77
& 0.82 / 0.83
& 0.90 / 0.89 \\

AF-Whisper
& 0.65 / 0.64
& 0.80 / 0.76
& 0.77 / 0.77
& 0.82 / 0.81
& 0.90 / 0.90 \\
\hline
\multicolumn{6}{c}{(continued)} \\
\hline
\textbf{Encoder}
& \textbf{BC19}
& \textbf{NISQA-LIVETALK}
& \textbf{NISQA-P501}
& \textbf{CHiME-7}
& \textbf{URGENT25-SQA} \\
\hline
WavLM
& 0.81 / 0.77
& 0.83 / 0.81
& 0.93 / 0.93
& 0.76 / 0.75
& 0.74 / 0.74 \\

Kimi-Audio
& 0.86 / 0.83
& 0.88 / 0.87
& 0.92 / 0.92
& 0.77 / 0.76
& 0.73 / 0.73 \\

Qwen3Omni
& 0.62 / 0.46
& 0.85 / 0.84
& 0.92 / 0.92
& 0.78 / 0.78
& 0.77 / 0.76 \\

AF-Whisper
& 0.72 / 0.59
& 0.84 / 0.82
& 0.89 / 0.91
& 0.78 / 0.78
& 0.72 / 0.72 \\
\hline
\end{tabular}
\label{tab:encoder_eval_corr_results_appendix}
\end{table*}

\begin{table*}[h]
\centering
\caption{Correlation (LCC / SRCC) between predicted and ground-truth quality metrics from F1C5M15-$\mathcal{D}_{ref}$ on URGENT-SQA test sets (excluding MOS). }
\setlength{\tabcolsep}{8pt}
\renewcommand{\arraystretch}{1.1}
\begin{tabular}{l c c}
\hline
\textbf{Metric} & \textbf{URGENT2024-SQA} & \textbf{URGENT2025-SQA} \\
\hline
UTMOS         & 0.99 / 0.98 & 0.98 / 0.98 \\
Distill-MOS  & 0.93 / 0.92 & 0.93 / 0.93 \\
SCOREQ       & 0.95 / 0.96 & 0.95 / 0.95 \\
NISQA-MOS    & 0.90 / 0.90 & 0.87 / 0.86 \\
\hline
DNSMOS        & 0.91 / 0.87 & 0.89 / 0.88 \\
LSD           & 0.45 / 0.35 & 0.48 / 0.47 \\
SDR           & 0.81 / 0.88 & 0.70 / 0.77 \\
PESQ          & 0.85 / 0.86 & 0.85 / 0.84 \\
PESQc2        & 0.90 / 0.90 & 0.86 / 0.86 \\
\hline
LPS            & 0.92 / 0.91 & 0.87 / 0.89 \\
SpeechBERTScore & 0.91 / 0.92 & 0.89 / 0.89 \\
ESTOI          & 0.87 / 0.91 & 0.89 / 0.90 \\
\hline
SpeakerSim    & 0.72 / 0.73 & 0.76 / 0.76 \\
\hline
MCD           & 0.73 / 0.66 & 0.68 / 0.69 \\
\hline
\end{tabular}
\label{tab:metric_corr_nonmos}
\end{table*}

\begin{figure*}[h] 
    \centering
    \includegraphics[width=.9\linewidth]{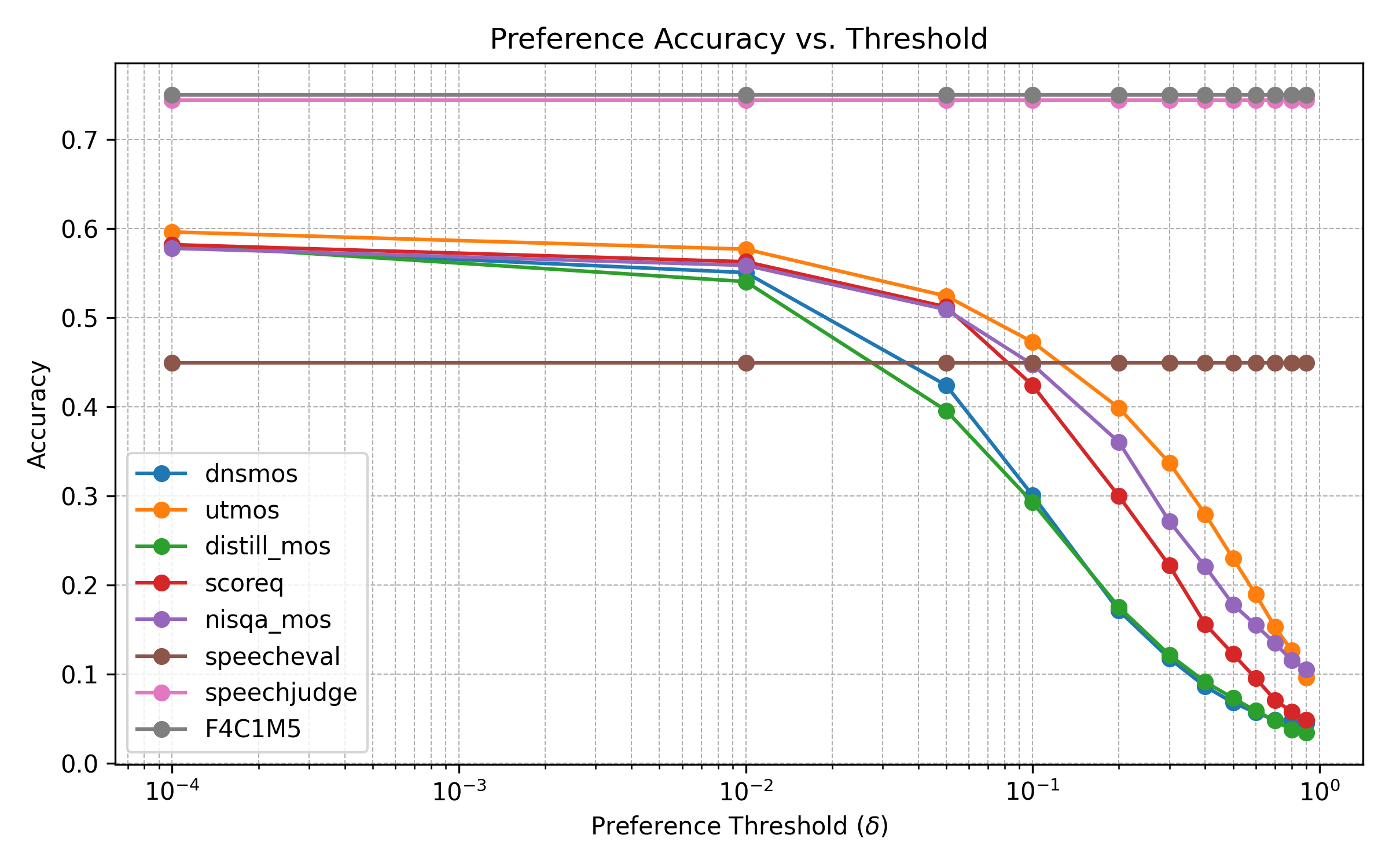} 
    \caption{Preference accuracy on the SpeechJudge test set versus tie threshold~\(\delta\).}
    \label{im:speechjudge_preference_accuracy_vs_threshold}
\end{figure*}

\section{Additional Evaluation Results}
\label{appendix:additional_evaluation_results}

Additional evaluation results for preference-based, correlation-based metrics and feature extractors are provided in Tables~\ref{tab:eval_pair_results_appendix}, Table~\ref{tab:eval_corr_results_appendix}, and Table~\ref{tab:encoder_eval_corr_results_appendix} respectively.

\section{The Use of Large Language Models}
We acknowledge the use of LLM for assisting with tasks such as grammar correction, enhancing expression diversity, formatting tables, and debugging code. However, we emphasize that all core ideas and experiments are original contributions.

\end{document}